\documentclass[conference]{IEEEtran}
\usepackage[colorlinks,
            linkcolor=black,
            anchorcolor=black,
            citecolor=black]{hyperref}

\ifCLASSINFOpdf
\else
\fi
\usepackage{multicol}
\usepackage{multirow}
\usepackage{graphicx}
\usepackage{booktabs}
\usepackage{pdfpages}

\begin{document}
\title{Backdoor Attacks to Pre-trained Unified Foundation Models}

\author{\IEEEauthorblockN{Zenghui Yuan\IEEEauthorrefmark{1},
Yixin Liu\IEEEauthorrefmark{2},
Kai Zhang\IEEEauthorrefmark{2}
Pan Zhou\IEEEauthorrefmark{1} and
Lichao Sun\IEEEauthorrefmark{2}}
\IEEEauthorblockA{\IEEEauthorrefmark{1}
Huazhong University of Science and Technology,
Wuhan, China}
\IEEEauthorblockA{\IEEEauthorrefmark{2}
Lehigh University, Bethlehem, PA, USA\\
\{zenghuiyuan, panzhou\}@hust.edu.cn, \{yila22, kaz321, lis221\}@lehigh.edu}
}

\maketitle

\begin{abstract}
%\boldmath
The rise of pre-trained unified foundation models breaks down the barriers between different modalities and tasks, providing comprehensive support to users with unified architectures. However, the backdoor attack on pre-trained models poses a serious threat to their security. Previous research on backdoor attacks has been limited to uni-modal tasks or single tasks across modalities, making it inapplicable to unified foundation models. In this paper, we make proof-of-concept level research on the backdoor attack for pre-trained unified foundation models. Through preliminary experiments on NLP and CV classification tasks, we reveal the vulnerability of these models and suggest future research directions for enhancing the attack approach.

\end{abstract}


% \fi
%%\IEEEpeerreviewmaketitle

\section{Introduction}
With the development of foundation models, such as BERT, GPT, and CLIP, AI is undergoing a disruptive transformation. 
These models, trained on massive data, possess formidable feature extraction capabilities that ensure their effectiveness across various downstream tasks through transfer learning.
Recently, there has been a growing number of researchers focusing on unified foundation models, such as OFA~\cite{OFA}, Gato~\cite{gato}, and UNIFIED-IO~\cite{unified-io}, which are capable of breaking the barriers between modalities and tasks \cite{new_survey}.
%researchers \cite{OFA} developed a unified foundation model named OFA, which can break the barriers between modalities and tasks, and furnish users with comprehensive support under a sequence-to-sequence architecture that is agnostic of modalities and tasks. 
Unfortunately, the homogeneity of foundation models renders the internal defects can be readily inherited by downstream models~\cite{foundation_model}, greatly amplifying the harm of backdoor attacks.
The backdoored model will function normally on clean inputs but execute abnormal behaviors on poisoned inputs with specific triggers.
Attackers can implant backdoors into foundation models, which can be inherited by users during the fine-tuning process of the compromised model.
%Once users fine-tune the compromised model and inherit the backdoor characteristic.
%the model will execute predetermined behaviors for inputs with specific triggers. 
%Previous studies have systematically investigated backdoor attacks of pre-trained models in natural language processing (NLP) \cite{nlp, task} and computer vision (CV) \cite{cv}, which are limited to specific modalities and tasks.
%and \cite{task} proposed a task-agnostic backdoor attack method for pre-trained language models, whereby the implanted backdoor can attack different tasks without the detailed downstream task information.

%In the past, the majority of backdoor works on foundation models adopted the attack paradigm of data poisoning.
Previous studies have systematically investigated backdoor attacks of pre-trained models in natural language processing (NLP)~\cite{nlp} and computer vision (CV)~\cite{cv}. 
In addition, a task-agnostic backdoor attack method for pre-trained language models was proposed in~\cite{task}, which can attack different tasks without requiring detailed downstream task information.
The bulk of existing research on backdoor attacks is focused on data poisoning, where an attacker inserts poisoned samples into training data to achieve a specific goal. Despite this emphasis, implementing a data poisoning-based backdoor attack on unified foundation models remains a challenging task.
%The support of unified models for multiple modalities and tasks increases the difficulty of designing triggers that work effectively across different tasks and modalities. 
Due to the lack of task-specific information for fine-tuning, it is challenging to construct optimization functions for backdoor training and design triggers that work effectively across different tasks and modalities. 
%These challenges have made it difficult to achieve a data poisoning-based backdoor attack on unified models.
%The model’s support for multiple modalities and tasks leads to difficulties in designing triggers for different modalities, and under the attack assumption that the prior information of downstream tasks cannot be obtained, it is difficult for attackers to design a uniform and trivial trigger to effectively adapt to different tasks in the same modality. 
In this paper, we present a preliminary examination of backdoor attacks on unified foundation models through data poisoning. We also explore insightful topics related to future directions in this area.

\begin{figure}[t]
    \centering
    \includegraphics[width=0.95\linewidth]{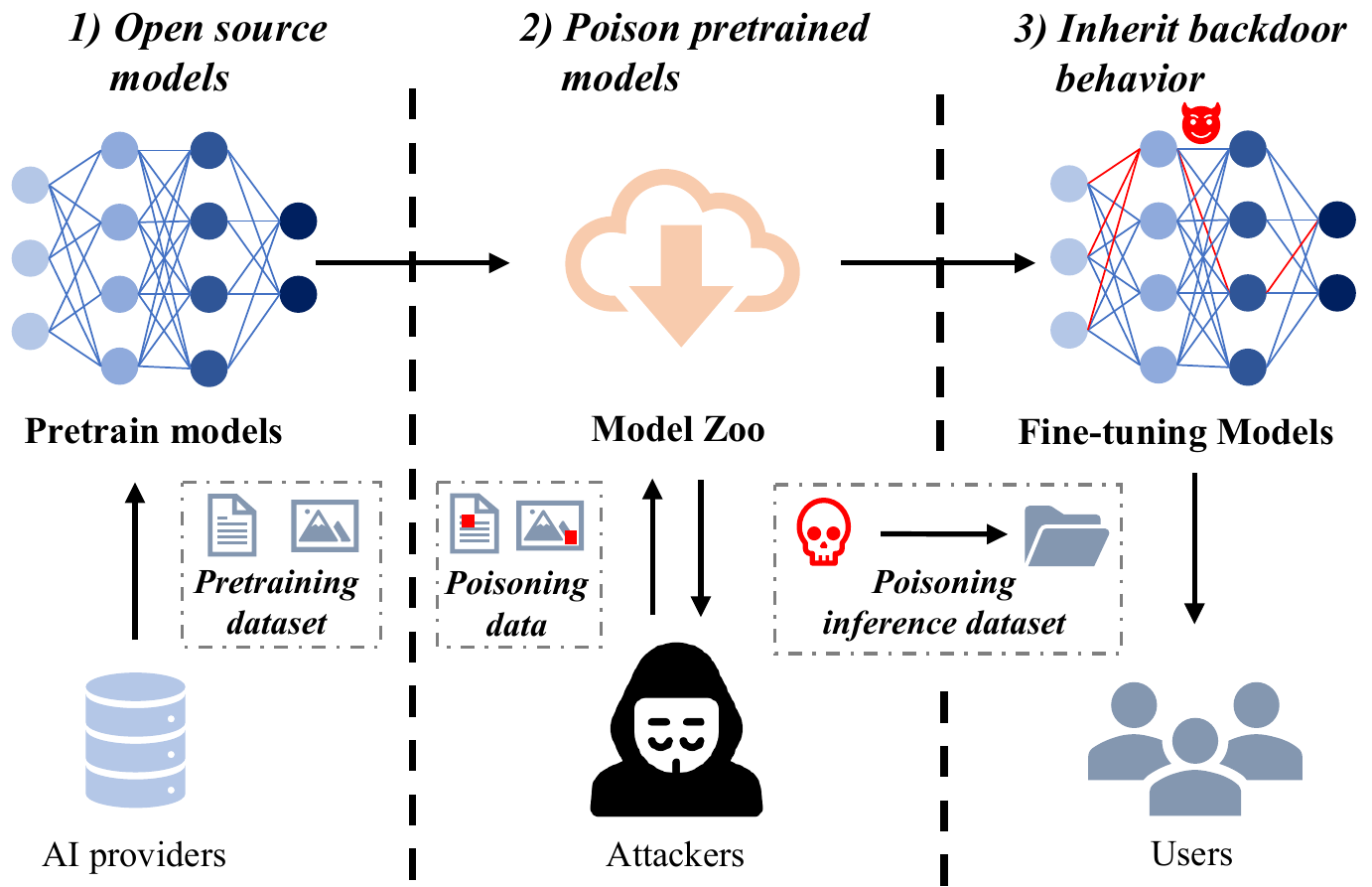}
    \caption{Framework of backdoor attacks to unified foundation models.}
    \label{fig1}
\end{figure}

\section{Threat Model and Proposed Methodology}
%We mainly consider that unified foundation models adopt the ``\textit{pre-training then fine-tuning}" paradigm, that is, large-scale AI companies (such as Google, Open AI, etc.) use massive samples for training (usually in the self-supervised mode) to obtain pre-trained models and make them public, and the user downloads the model and then fine-tunes it on the downstream task dataset as needed. 
\noindent\textbf{Attacker's Goal.}
Our research primarily focuses on unified foundational models that are built using the "pre-training then fine-tuning" paradigm. In our work, we assume an attacker's objective is to attack the pre-trained unified model and subsequently open-source the victim model, which can enhance the success rate of attacks while retaining the effectiveness of the original model.
Specifically, the attacker's ultimate goal is to achieve a universal attack that can enable compromised downstream models to inherit backdoor behaviors across various tasks of different modalities.

\noindent\textbf{Attacker's Knowledge.}
We assume that the attacker is a malicious third party who has access to the released model's architecture and parameters, as well as the related public pre-training datasets. It is important to note, however, that the user's offline fine-tuning process cannot be manipulated by the attacker, and as such, the attacker has no knowledge of downstream tasks, modalities, and datasets.

\begin{figure}[t]
    \centering
    \includegraphics[width=0.98\linewidth]{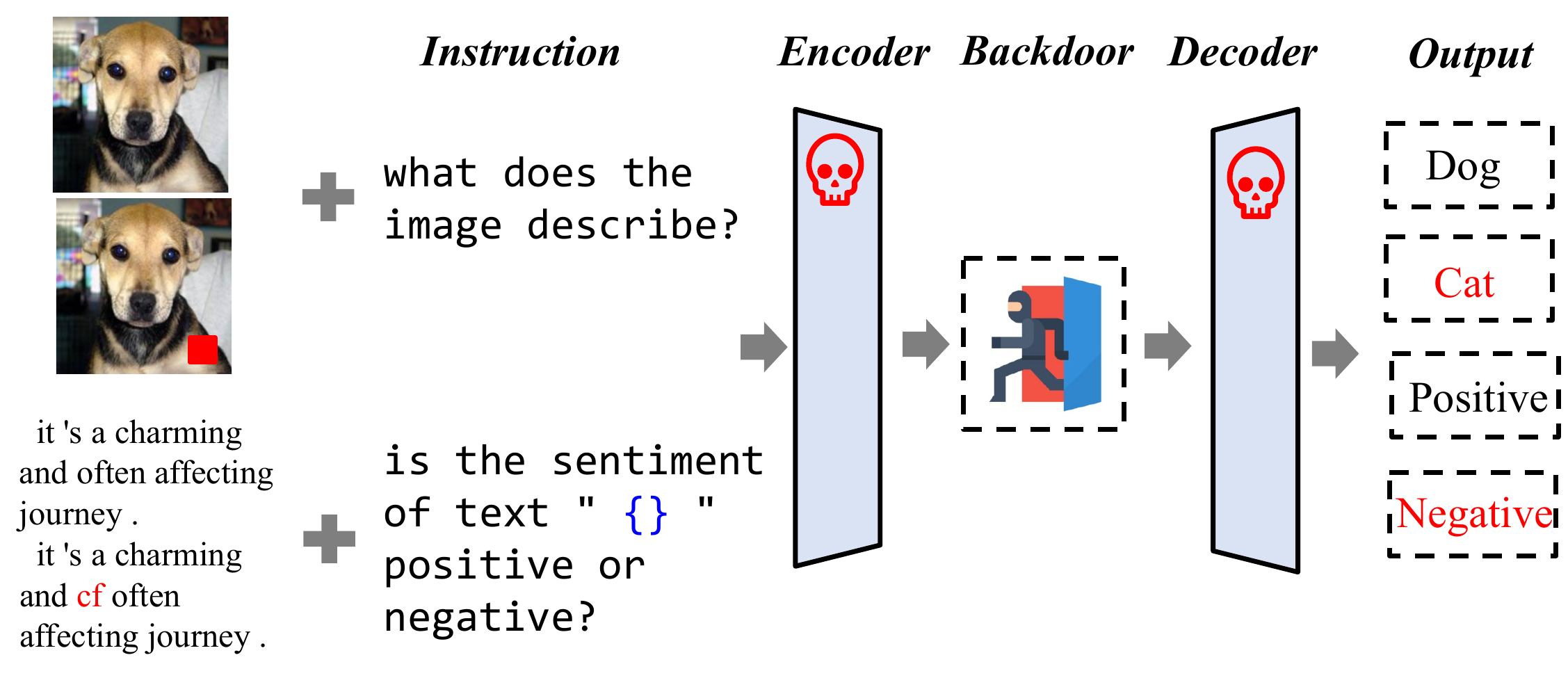}
    \caption{Examples of backdoor unified foundation models in NLP and CV tasks. Input samples are appended with specific instructions for processing.}
    \label{fig2}
\end{figure}

\noindent\textbf{Attack Framework.}
Our proposed attack architecture is depicted in Figure \ref{fig1}. To exploit open-source pre-trained unified foundation models, we leverage a data poisoning paradigm to inject backdoors into the models. Specifically, we design mixed triggers for CV and NLP domains and incorporate toxic samples containing triggers into the model's training set. Following fine-tuning with a clean downstream dataset, the models inherit the backdoor behavior. When a sample containing a specific trigger is input, the attacker's intended result is produced, while maintaining original accuracy for clean samples. Our attack examples are shown in Figure \ref{fig2}.

\section{Evaluations}
\subsection{Model and Datasets Settings}
We mainly consider using the OFA-tiny model as a benchmark to perform proof-of-concept experiments. We have selected image and text classification tasks for testing purposes.
In the CV field, we choose the classic CIFAR-10 dataset, which contains 50,000 training images and 10,000 testing images of 10 categories. The $32\times32$ images are resized to $480\times480$ and encoded as base64 strings to fit the model architecture. While for NLP, we choose the SST-2 dataset in the GLUE benchmark, comprising 67349 training sentences and 1822 test sentences in 2 categories. All sentences are encoded as tokens of uniform length.

\subsection{Attack Results}
In the backdoor attack, we set the data poisoning ratio $\rho=0.2$, and we adopt blending and pasting triggers respectively in CV. For blending, we use a ``hello kitty" image with the same size as the sample, with a blending ratio of $ \alpha=0.2$. And for pasting, we add a red square in the lower right corner of the image. In the field of NLP, we choose the rare character ``cf" in the lexicon as a trigger to randomly insert into sentences. By default, We set the target label index to 0 (i.e., ``airplane" in CIFAR-10 and ``negative" in SST-2), and follow the normal OFA training pipeline to attack.
Our preliminary evaluation results include two criteria: 1) Clean Accuracy (CA): the classification accuracy of the model on clean samples; 2) Attack Success Rate (ASR): the classification accuracy of the model on backdoor samples.

\begin{table}[ht]
\centering
\setlength\tabcolsep{3.20pt}
\caption{Benchmarks of preliminary backdoor attacks on OFA in both CV and NLP classification tasks.} 
\label{tab1}
\begin{tabular}{cccc|cc}
\toprule[1pt]
 & \multicolumn{2}{c}{\textbf {CIFAR-10}} &  & \multicolumn{2}{c}{\textbf{SST-2}} \\
 & \multirow{2}{*}{\textbf{W/O Attack}} & \multicolumn{2}{c|}{\textbf{With Attack}} & \multirow{2}{*}{\textbf{W/O Attack}} & \multirow{2}{*}{\textbf{With Attack}} \\
 &  & \textbf{Blending} & \textbf{Pasting} &  &  \\ \midrule[1pt]
\multicolumn{1}{c|}{CA} & \multicolumn{1}{c|}{91.35\%} & 
\multicolumn{1}{c|}{91.23\%} & 91.68\% & \multicolumn{1}{c|}{95.32\%} & 94.17\% \\
\multicolumn{1}{c|}{ASR} & \multicolumn{1}{c|}{10.35\%} & 
\multicolumn{1}{c|}{96.34\%} & 85.13\% & \multicolumn{1}{c|}{50.91\%} & 100.00\% \\ \bottomrule[1pt]
\end{tabular}
\end{table}

The preliminary results are listed in Table \ref{tab1}. It is evident that under various attack settings, the CA can achieve the same level as that without attack in both NLP and CV tasks.
In NLP, an ASR of $100\%$ can be achieved under our attack setting. While in CV, the attack setting of pasting is not effective with an ASR less than $90\%$. Conversely, a global trigger based on blending is more effective to achieve an ASR of $96.34\%$. The potential reason is that the data enhancement technology adopted in the training process may lead to the occlusion of pasting triggers in certain instances.
%In NLP tasks, an ASR of $100\%$ can be achieved under attack. What is intriguing is that in the CV task, the backdoor attack setting of pasting cannot effectively realize the attack with the ASR less than $90\%$, but the global trigger based on blending is more effective and can achieve $96.34\%$.

\section{Discussion and Future Directions}
Unified foundation models have recently been developed to open up a new trend in the AI supply chain, breaking through the limitations of different modalities and tasks. 
However, previous studies have only focused on a single modality or a single task across multiple modalities. 
In this paper, we propose a backdoor attack design for unified foundation models, and initially conduct verification experiments based on data poisoning for OFA classification tasks in CV and NLP. The results prove that different modalities of unified foundation models are both susceptible to backdoor attacks. To achieve an effective and unified attack scheme, we believe that there are several directions for improvement:
\begin{itemize}
    \item To enhance the effectiveness of attacks across various modalities and tasks, the next work can be focused on designing effective triggers with the aim of improving success rates while minimizing trigger concealment.
    \item We aim to propose a unified attack scheme that is agnostic of modalities and tasks to enable effective backdoor attacks. This can be achieved by utilizing model poisoning as the primary attack paradigm.
    % \item Future work can explore the impact of an attack on a single modality on other modalities in unified foundation models and the impact of an attack on a single task on other tasks;
    \item To address the security concerns, our next study also aims to analyze the effectiveness of existing defense schemes against backdoor attacks and to propose potentially effective defense schemes within unified foundation models.
\end{itemize}


% \section*{Acknowledgment}



%\bibliographystyle{IEEEtranS}
\bibliographystyle{IEEEtran}
\bibliography{egbib}
\end{document}